\documentclass[aps,prd,twocolumn,groupedaddress,nofootinbib,showpacs]{revtex4}

\usepackage{amssymb}
\usepackage{amsbsy}
\usepackage{amsmath}

\newcommand{\rf}[1]{(\ref{#1})}
\newcommand{\beq}{\begin{equation}}
\newcommand{\eeq}{\end{equation}}
\newcommand{\be}{\begin{equation}}
\newcommand{\ee}{\end{equation}}
\newcommand{\bea}{\begin{eqnarray}}
\newcommand{\eea}{\end{eqnarray}}
\newcommand{\eq}[1]{Eq.~(\ref{#1})}
\newcommand{\non}{\nonumber \\*}
\newcommand{\ie}{{i.e.}\ }

\newcommand{\e}{\,\mbox{e}}
\renewcommand{\d}{{\rm d}}
\renewcommand{\i}{{\rm i}}
\newcommand{\blambda}{\bar\lambda}
\newcommand{\brho}{\bar\rho}

\newcommand{\A}{A}
\renewcommand{\t}{\tau}
\newcommand{\s}{\sigma}

\newcommand{\eps}{\varepsilon}

\newcommand{\tr}{\mathrm{tr}}
\newcommand{\LA}{\left\langle}
\newcommand{\RA}{\right\rangle}

\def\fun#1#2{\lower3.6pt\vbox{\baselineskip0pt\lineskip.9pt
\ialign{$\mathsurround=0pt#1\hfil##\hfil$\crcr#2\crcr\sim\crcr}}}

\begin{document}


\title{Scattering amplitudes of regularized bosonic strings}

\author
{J. Ambj\o rn$\,^{a,b}$ and Y. Makeenko$\,^{a,c}$}

\affiliation{\vspace*{2mm}
${}^a$\/The Niels Bohr Institute, Copenhagen University,
Blegdamsvej 17, DK-2100 Copenhagen, Denmark\\
${}^b$\/IMAPP, Radboud University, Heyendaalseweg 135,
6525 AJ, Nijmegen, The Netherlands\\
${}^c$\/Institute of Theoretical and Experimental Physics,
B. Cheremushkinskaya 25, 117218 Moscow, Russia\\
	\vspace*{1mm}
{email: ambjorn@nbi.dk \ makeenko@nbi.dk}
}


\begin{abstract}
We compute scattering amplitudes of the regularized bosonic Nambu-Goto
string in the mean-field approximation, disregarding fluctuations
of the Lagrange multiplier and an independent metric about their 
mean values.  We use the previously introduced Lilliputian scaling limit
to recover  the Regge behavior of the amplitudes
with the usual linear Regge trajectory in space-time dimensions $d>2$.
We  demonstrate 
a stability of this minimum of the effective action 
under fluctuations for $d<26$.
\end{abstract}

\pacs{11.25.Pm, 11.15.Pg,} 

\maketitle

\section{Introduction}

String theory emerged in very early 1970's from dual resonance models which were
introduced to explain linear Regge trajectories of hadrons.
Canonical quantization of relativistic bosonic string is consistent only in $d=26$ 
dimensions and on mass shell. 
These restrictions can be potentially overcome by an alternative path-integral string quantization \cite{Pol81}, 
where the problem remains nonlinear in $d\neq 26$ and/or off mass shell
even after a gauge fixing.
More precisely, these nonlinearities are due to the fact that the ultraviolet cutoff 
depends on 
the metric on the string world sheet, as is prescribed by diffeomorphism invariance. 
They are natural when one uses the proper time regularization, but they are 
hardly seen for the zeta-function
regularization which is intimately linked to the mode expansion used in 
canonical quantization.

In the recent papers \cite{AM15b,AM17} we have analyzed the bosonic Nambu-Goto string
in the mean field approximation, where the world sheet metric  can be  substituted by its mean value.
This approximation becomes exact at large $d$ and is applicable for finite $d$. We considered
the string with the Dirichlet boundary conditions and computed its ground state energy as a function of the string length. We showed that the results obtained by  
canonical quantization are reproduced if the bare string tension $K_0$ approaches 
its critical value $K_*$ from above:
\be
K_0\to K_*+ \frac {K_R^2}{2\Lambda^2\sqrt{d^2-2d}},~~
K_*=\left(d-1+\sqrt{d^2-2d}\right)\! \Lambda^2\!,
\label{sca}
\ee
where $K_R$ stands for the renormalized string tension.
Associated with this scaling limit was a renormalization of length scales~\cite{AM15b,AM17}
\be
L_{R} = \sqrt{ \frac {d+\sqrt{d^2-2d}}{2K_R}}\,\Lambda L_0.
\label{jlph}
\ee
In the limit where the cutoff $\Lambda \to \infty$ and $K_R$ and $L_{R}$ stay 
finite, the ground state energy also stays finite and
agrees with the one obtained using canonical quantization.
Such a limit is possible because the bare metric on the worldsheet becomes singular.
We called the scaling limit ``Lilliputian'' since the 
bare  length scales   as $L_0 \sim 1/{\Lambda}$.

The goal of this Paper is to further understand 
the Lilliputian string world and, in
particular, the meaning of the performed renormalization of the length scale by computing
scattering amplitudes.

\section{Nambu-Goto string in the mean-field approximation\label{s:NG}}

Our starting point is the Nambu-Goto string whose action is rewritten in the standard way by
introducing a Lagrange multiplier $\lambda^{ab}$ and 
an independent metric  $\rho_{ab}$
as
\bea
S_{\rm NG}&=&
K_0 \!\int \d^2z\,\sqrt{\det \partial_a X \cdot \partial_bX} 
=
K_0 \!\int \d^2z\,\sqrt{ \det\rho_{ab}}\non &&
+\frac{K_0}2 \!\int \d^2z\, \lambda^{ab} \left( \partial_a X \cdot \partial_bX -\rho_{ab}
\right).
\label{aux}
\eea
The path integration goes independently  over real values of $X^\mu$ and $\rho_{ab}$
and imaginary values of $\lambda^{ab}$. To obtain the effective action
governing the fields $\lambda^{ab}$ and $\rho_{ab}$, we
split $X^\mu=X^\mu_{\rm cl} +X^\mu_{\rm q}$ and perform the Gaussian path
integral over $X^\mu_{\rm q}$. Fixing the conformal gauge $\rho_{ab}=\rho \delta_{ab}$,
we find
\bea
\lefteqn{S_{\rm eff}= K_0 \int \d^2 z\,{\rho} } \non &&\!
+\frac{K_0}2 \int \d^2 z\, \lambda^{ab} \left( \partial_a X_{\rm cl } \cdot \partial_bX_{\rm cl } -\rho \delta_{ab}
\right)  \non &&\!
+ \frac d2 \tr \log\! \left[ -\frac1 {{{\rho}} }  \partial _a \lambda^{ab} \partial_b  \right]\! -\tr \log 
\!\left[ -\frac1 {{{\rho}} }  \partial ^2  +\frac1 {2{{\rho}} }  (\partial^2  \log \rho)
\right]\! ,
\non
\label{aux1}
\eea
where the last term on the right-hand side comes from the ghost determinant.

For definiteness we can use the proper-time regularization of the traces in \eq{aux1}
\be
\tr\log {\cal O} 
= - \int_{a^2}^\infty \frac{\d\tau}\tau \int \d^2 \omega  
\LA \omega |\e^{-\tau {\cal O}} |\omega \RA,
\label{5}
\ee
where the proper-time cutoff $a^2$ is related to the momentum cutoff $\Lambda^2$ by
\be
a^2= \frac1{4\pi \Lambda^2 },
\label{aLa}
\ee
but the results will not depend on the regularization used.

In the mean-field approximation, which becomes exact at large $d$, 
we can disregard fluctuations
of  $\lambda^{ab}$ and $\rho$ about their saddle-point values,
\ie simply substitute them by the mean values. 
This is analogous to the study
of the $N$-component sigma-model at large $N$, where we can disregard quantum
fluctuations of the Lagrange multiplier.

\section{Mean field for scattering amplitude}

To compute the scattering amplitude, we introduce a piecewise constant momentum loop
\be
P (\t)= \sum _k p_k \theta\left( \t-\t_k\right),\quad \sum_k p_k=0
\label{piecewise}
\ee
and consider
\be
{\cal A}\left[P(\cdot)\right]=\LA \e^{\i \int P_\mu \d x^\mu} \RA
=\int {\cal D}X^\mu \,\e^{-S_{\rm NG} +\i \int P_\mu \d x^\mu},
\label{amplitude}
\ee
where $x^\mu(\t)$ is the value of $X^\mu$ at the boundary modulo a reparametrization
of the boundary. 
For the upper half-plane coordinates $z=\s+ \i y$ the boundary is along the real axis, $y=0$,
and we have explicitly $X^\mu(\s,0)=x^\mu(\t(\s)) $ with the 
nonnegative derivative $\d \t /\d \s\geq 0$.

Integrating by parts in the exponent in \eq{amplitude} we obtain
\be
\int P_\mu \d x^\mu = -\sum_k p_k \cdot x_k,\quad x_k=x(\t_k),
\ee
which reproduces the string vertex operators for scalars.
The averaging in \eq{amplitude} can be also represented  as
the path integral over $\lambda^{ab}$ and $\rho$
with the effective action \rf{aux1}.
We shall return to this issue in Sect.~\ref{s:metr}.

Let us regularize \rf{piecewise} as
\be
P (\t)=\frac1\pi  \sum _k p_k \arctan\left(\frac{\t-\t_k}{\sqrt{\eps_k}}\right),
\quad \sum_k p_k=0,
\label{reguloop}
\ee
where
\be
\eps_k=\eps \frac{(\tau_{k+1}-\tau_k)(\tau_k-\tau_{k-1})}{(\tau_{k+1}-\tau_{k-1})}
\label{epsis}
\ee
to comply with diffeomorphism invariance.
Then it becomes clear that \rf{reguloop}  is a singular parametrization of a polygonal momentum loop. This is like the Wilson-loop/scattering-amplitude duality in the ${\cal N}=4$
supersymmetric Yang-Mills theory \cite{AM07} which was extended to QCD string in \cite{MO08}.

For the scattering amplitude it is convenient to use the upper half-plane coordinates, where
the boundary of the string world sheet is parametrized by the real axis, resulting in 
the Koba-Nielsen variables. 
Repeating the technique of Sect.~\ref{s:NG} for integrating over $X^\mu$, now with the Neumann boundary conditions, we obtain for constant $\blambda$ 
(to be justified below) the effective action 
\bea
S_{\rm eff}&=&\frac{1}{2K_0\blambda} \sum_{i,j} p_i G_\eps\left(
\frac{\s_i-\s_j}{\sqrt{\blambda}}\right) p_j
+{K_0} \A \left( 1-\blambda  \right)
 \non &&
+\frac d2 \tr \log\!\left [-\frac{\blambda}\brho
\partial ^2 \right]\! -\tr \log\!\left [-\frac1\brho
\partial ^2+\frac1{2\brho} (\partial^2 \log\brho)\right]\!. \non &&
\label{SeA}
\eea
Here  we have denoted $\s_i=\s(\t_i)$,
$G_\eps$ is a regularized Green function of the type
\be
G_\eps(\s)=-\frac1{2\pi} \log \left( \s^2+\eps(\s)\right),
\ee
and we have used the notation
\be
\A=\int \d^2 z \,\brho  .
\ee

For the $2\to 2$ amplitude we have four $\s_k$'s ($k=1,\ldots,4$), but the 
amplitude depends only on the projective-invariant ratio
\be
r=\frac{(\s_4-\s_3)(\s_2-\s_1)}{(\s_4-\s_2)(\s_3-\s_1)}.
\ee
{}From the conformal mapping of the upper half-plane 
onto a $\omega_L\times \omega_R$ rectangle we have
\be
\frac{\omega_L}{\omega_R}=\frac{K\left(\sqrt{r}\right)}{K\left(\sqrt{1-r}\right)},
\ee
where $K$ is the complete elliptic integral.

Noting that
\bea
\sum_{i,j=1}^4 p_i G_\eps\left(
\frac{\s_i-\s_j}{\sqrt{\blambda}}\right) p_j &=&- s\log r -t \log (1-r) 
\non &&-\frac{ \log \eps}{2\pi}
\sum_{i=1}^4 p_i^2
\label{diver}
\eea
with $s=-(p_1+p_2)^2$, $ t=- (p_1+p_4)^2$ being Mandelstam's
variables and dropping the last term on the right-hand side of \eq{diver}
for lightlike momenta $p_i^2=0$, 
we find
\bea
S_{\rm eff}&=&\frac{1}{2\pi K_0\blambda} 
\left[ s\log r +t \log (1-r)\right]
+{K_0} \A \left(1-\blambda \right) \non && -\frac{d \Lambda^2 A}{2\blambda} +
{\Lambda^2 A}
+\frac {d-2}{24} \log\left[r(1-r) \right] ,
\label{187}
\eea
while the (ordered) integration over $\s_i$'s 
is inherited from reparame\-triza\-tions of the boundary. 
It gives the volume of the projective group times 
the integration over $r$, which is equivalent to the integration over the ratio
$\omega_R/\omega_L$.
The boundary terms are negligible at least for lightlike momenta as we show below.

It is worth noting that our procedure of the smearing \rf{reguloop} is similar
to the one introduced in \cite{DOP84} for computing the L\"uscher term for
a rectangular Wilson loop. Actually, the last term in \eq{187}, coming from the determinants, for $s\gg t$ makes sense of the momentum 
L\"uscher term~\cite{Jan01,Mak11a}.
However, it 
is {\em exact}\/ for arbitrary $r$ in view of the identity~\cite{AMS14}
\bea
\lefteqn{\frac1{2\left(K\left( \sqrt{1-r}\right) \right)^{1/2}} \,
\eta \left( i \frac{K\left( \sqrt{1-r}\right) }{K\left( \sqrt{r}\right) }  
 \right)} \non &&\hspace*{1cm}= 
\frac{1}{2^{5/6} \pi^{1/2} }  \left[r(1-r)\right]^{1/12}
\label{ide}
\eea
for the  Dedekind $\eta$-function:
\be
\prod_{m,n=1}^\infty \left( \frac{\pi m^2}{\omega_R^2}+ 
\frac{\pi n^2}{\omega_L^2} \right) = \frac1{\sqrt{2\omega_L} }\,
\eta \left(i \frac{\omega_R}{\omega_L}   \right).
\label{180}
\ee

Minimizing \rf{187} with respect to $A$, we find
\be
\blambda =\frac12+\frac{\Lambda^2}{2K_0}+\sqrt{\frac14\left(1+\frac{\Lambda^2}{K_0}\right)^2-\frac{d\Lambda^2}{2K_0}}
\ee
which is the same value as found in \cite{AM17}. 
Minimizing \rf{187} with respect to $\blambda$, we obtain
\be
\A= \frac{1}{2\pi K_0^2}\frac{1}{(2\blambda-1-\Lambda^2/K_0)\blambda}
\left[ s\log r +t \log (1-r)\right].
\label{22}
\ee
For the saddle-point value of the effective action we thus have
\bea
S_{\rm eff}&= &\left(\frac1{2\pi K_0 \blambda} s+\frac {d-2}{24} \right) 
\log r \non &&
+\left(\frac1{2\pi K_0 \blambda} t+\frac {d-2}{24} \right) 
\log (1-r).
\label{75}
\eea

If we integrate $\e^{-S_{\rm eff}}$ over $r$ as is prescribed by the path 
integral over reparame\-triza\-tions, we get the Veneziano amplitude 
${\cal B}\left(-\alpha(s),-\alpha(t)\right)$ with
\be
\alpha(s)=\frac1{2\pi K_0 \blambda} s+\frac {d-2}{24}, 
\quad \alpha(t)=\frac1{2\pi K_0 \blambda} t+\frac {d-2}{24}
\label{Rtra}
\ee
to {\em all orders}, not only semiclassically  like in textbooks.
This is a remarkable consequence of the identity \rf{ide}.

Minimizing \rf{75} with respect to $r$, we find
\be
r_*= \frac {\alpha(s)}{\alpha(s)+\alpha(t) } 
\label{r*}
\ee
and
\be
S_{\rm eff}= \alpha(s) 
\log \frac {\alpha(s)}{\alpha(s)+\alpha(t)} 
+\alpha(t) 
\log \frac {\alpha(t)}{\alpha(s)+\alpha( t) } .
\ee
For $s\gg t $ this results in the Regge behavior 
\be
{\cal A} \sim s^{\alpha(t)}
\ee
with the linear Regge trajectory \rf{Rtra}.

\section{Scaling limit and renormalization}

As mentioned in the introduction, the Lilliputian scaling regime was previously defined 
for a string with the Dirichlet boundary conditions by the following  
renormalization of string tension and the length scale~\cite{AM15b,AM17}
\be
K_{R}= K_0 
{\sqrt{\left( 1+\frac{\Lambda^2}{K_0}  \right)^2-\frac{2d\Lambda^2}{K_0}}}
=K_0 \left(2\blambda-1-\frac {\Lambda^2}{K_0} \right),
\label{KR}
\ee  
\be
L_{R} = \sqrt{\frac {\blambda}{2\blambda -1 -\frac{\Lambda^2}{K_0}}} \;
L_0.
\label{lph}
\ee
Equations
\rf{KR} and \rf{lph} are just slight rewritings of \rf{sca} and \rf{jlph} and the scaling 
limit is $\Lambda \to \infty$ while $L_R$ and $K_R$ stay finite.

Motivated by the length-scale renormalization \rf{lph}, we write
\be
s=\frac{\blambda }{2\blambda-1-\frac{\Lambda^2}{K_0}} s_{R},\quad
t=\frac{\blambda }{2\blambda-1-\frac{\Lambda^2}{K_0}} t_{R}
\label{193}
\ee
in the scaling regime \rf{sca}. Accounting for the renormalization of the string tension
\rf{KR}, we have  
\bea
S_{\rm eff}&=&\left( \frac{1}{2\pi K_{R} }s_{R}+\frac {d-2}{24}
\right)\log r  \non &&+ \left( \frac{1}{2\pi K_{R} }t_{R}+\frac {d-2}{24} \right)\log (1-r),
\eea
resulting in the linear renormalized Regge trajectory
\be
\alpha(t)=\alpha'_{R}t_{R}+\frac {d-2}{24},\quad
\alpha'_{R}=\frac{1}{2\pi K_{R} }
\ee
with a finite slope $\alpha'_{R}$.

\section{Metric and the boundary term\label{s:metr}}

Equation~\rf{22} represents the mean area of fluctuating surfaces, while for 
the computation of the boundary term in the effective action we need
the metric $\brho(z)$ itself. It can be computed as an average of the induced metric
\be
\brho_{ab}(z)= \LA
\partial _a X(z) \cdot  \partial _b X(z) \RA.
\label{treea}
\ee
To understand the structure of the boundary term, we shall compute \rf{treea}
at the tree level, \ie in the classical approximation.

The harmonic function in the upper half-plane with the boundary conditions \rf{reguloop}
at the real axis is 
\be
P (x,y)=\frac1\pi  \sum _k p_k \arctan\left(\frac{ x-\s_k}{y+\sqrt{\epsilon_k}}\right),
\quad \sum_k p_k=0.
\label{regulooph}
\ee
By T-duality the computation is the same as for for the world sheet
\be
X^\mu_{\rm cl} =\frac 1{K_0\blambda} P^\mu.
\label{34}
\ee

Using \rf{regulooph}, \rf{34} we obtain for the classical induced metric
\begin{widetext}
\begin{subequations}
\bea
\partial _1 X_{\rm cl} \cdot  \partial _1 X_{\rm cl} 
&=&\frac1{\pi^2 K_0^2\blambda^2} \sum _{i,j}
\frac{p_i\cdot p_j \,(y+\sqrt{\varepsilon_i})(y+\sqrt{\varepsilon_j})}
{[(x-\s_i)^2+(y+\sqrt{\varepsilon_i})^2][(x-\s_j)^2+(y+\sqrt{\varepsilon_j})^2]},
\label{35} \\
\partial _1 X_{\rm cl} \cdot  \partial _2 X_{\rm cl} &=&
-\frac1{\pi^2 K_0^2 \blambda^2} \sum_{i,j} 
\frac{p_i \cdot  p_j\, (x-\s_i)(y+\sqrt{\varepsilon_j})}{[(x-\s_i)^2+(y+\sqrt{\varepsilon_i})^2][(x-\s_j)^2+(y+\sqrt{\eps_j})^2]},
\label{g12}\\
\partial _2 X_{\rm cl} \cdot  \partial _2 X_{\rm cl} 
&=&\frac1{\pi^2K_0^2\blambda^2} \sum_{i,j} 
\frac{ p_i \cdot  p_j\, (x-\s_i)(x-\s_j)}
{[(x-\s_i)^2+(y+\sqrt{\eps_i})^2][(x-s_j)^2+(y+\sqrt{\eps_j})^2]}.
\label{g22}
\eea
\label{ggg}
\end{subequations}
\end{widetext}
 As is shown in \cite{Mak11a},
\rf{g12} vanishes and  \rf{35} coincides with 
 \rf{g22}, provided  $\s_i$'s satisfy
\be
\sum_{j\neq i} \frac{p_i \cdot p_j}{\s_i-\s_j}=0
\label{scaeq}
\ee
for lightlike momenta. 
Then the metric  becomes conformal 
(\ie $\brho_{ab}=\brho\delta_{ab}$).

Equation~\rf{scaeq}
is the same condition as the recently advocated tree-level scattering equation~\cite{SA}.
Because of the projective symmetry three equations in \eq{scaeq} are not independent,
so for the case of four particles there is only one independent equation. This equation 
is nothing but the tree-level approximation of \eq{r*}. Notice that \eq{r*}
itself sums up all loops and guarantees that
$\brho_{ab}$ given by \rf{treea} is conformal   in the mean-field approximation.

The classical metric at the boundary 
\bea
\lefteqn{\partial _1 X_{\rm cl} \cdot  \partial _1 X_{\rm cl} 
\big|_B }
\non &&=\frac1{\pi^2 K_0^2\blambda^2} \sum _{i,j}
\frac{p_i\cdot p_j \,\sqrt{\varepsilon_i}\sqrt{\varepsilon_j}}
{\left[(x-\s_i)^2+\varepsilon_i\right]\left[(x-\s_j)^2+\varepsilon_j\right]}~~~
\eea
 vanishes except near the points $x=\s_k$ associated with edges of the polygon.
The integration along  the boundary
\be
\int _{-\infty}^{+\infty} \d x \sqrt{\partial _1 X_{\rm cl} \cdot  \partial _1 X_{\rm cl} 
\big|_B }=
\frac1{K_0\blambda} \sum_k \sqrt{p_k^2}
\label{39}
\ee
reproduces the length.
Thus the boundary term in the effective action is proportional to \rf{39}.
It vanishes for lightlike momenta $p_k^2=0$. Otherwise its contribution has to be
analyzed and remains finite in the scaling limit~\rf{sca}. 

We do not expect this boundary term (like the last term on the right-hand side of \eq{diver})
to have any effect on the Regge trajectory \rf{Rtra} because it depends on the masses
$\sqrt{p_k^2}$ of colliding particles, while the Regge trajectory does not.

To avoid the problem with the boundary terms, it is tempting to consider the case of closed
string scattering, when the boundary conditions are periodic along both axes
(the torus topology).
Then the boundary terms in the determinant are missing and what remains is four times larger than for the disk. This would change $(d-2)/24$ to 
$(d-2)/6$ in the above formulas as usual.

\section{Stability of fluctuations}

We can check if fluctuations about the mean values of $\lambda^{ab}$ and $\rho$
are stable in the quadratic approximation, following the considerations in
\cite{AM17}. The only difference from \cite{AM17} is that $\brho(z)$ is now  
coordinate-dependent, a dependence given by the right-hand side of \eq{35}.

Let us first consider the divergent part
of the effective action to quadratic order in fluctuations for the nondiagonal 
element of $\lambda^{ab}$. It is  given by Eq.~(25) of \cite{AM17}:
\bea
S_{\rm div}&=&\int\d^2 z
\left[ \frac{K_0}2 \lambda^{ab} \partial_a X_{\rm cl}\cdot \partial_b X_{\rm cl}+
K_0  \rho\Big(1 -\frac12  \lambda^{aa} \Big) \right.\non  && \left.-
\frac {d \Lambda^2}2   \frac \rho{\sqrt{\det{\lambda}}}
+ \Lambda^2   \rho \right], 
\quad\lambda^{aa}=\lambda^{11}+\lambda^{22}.
\label{cla}
\eea
Here $X_{\rm cl}$ in given by \eq{34} for the present case.

Expanding to quadratic order 
\bea
~~\sqrt{ \det(\blambda \delta^{ab}+\delta \lambda^{ab})}=\lambda+\frac12 \delta \lambda^{aa}
-\delta \lambda_2 
+{\cal O}\left((\delta \lambda)^3\right), \non
\delta \lambda_2 =\frac1{8\blambda} (\delta \lambda_{11}-\delta \lambda_{22})^2+ 
\frac 1{2 \blambda}(\delta \lambda_{12})^2,~~~~~~~~~~
\label{3001}
\eea
we find from \rf{cla} for constant $\blambda^{ab}=\blambda\delta^{ab}$
\bea
S^{(2)}_{\rm div} &= &-\frac{d\Lambda^2}{2\blambda} \!\int \d^2 z\, \brho\,\delta \lambda_2
-\left(K_0-\frac{d\Lambda^2}{2\blambda^2}\right)\!\int \d^2 z \, \delta \rho \frac{ \delta \lambda^{aa}}2 \non &&
-\frac{d\Lambda^2}{2\blambda^3}\!\int \d^2 z \, 
 \brho\left( \frac{ \delta \lambda^{aa}}2\right)^2 .
\label{Sdi}
\eea

The first term on the right-hand side of \eq{Sdi} plays a very important role for dynamics
of quadratic fluctuations. Because the path integral over $\lambda^{ab}$ goes
 parallel to imaginary axis,
\ie  $\delta \lambda^{ab}$ is pure imaginary, 
 the first term is always {\em positive}. 
Moreover, 
its exponential plays the role of a (functional) delta-function as $\Lambda\to\infty$,
forcing $\delta \lambda^{ab}=\delta \lambda \,\delta^{ab}$.

For $s\gg t $
we keep only the bulk term  to get
the effective action to quadratic order in fluctuations
\begin{widetext}
\bea
\delta  S_2&=&-\left(K_0-\frac{d\Lambda^2 }{2\blambda^2}\right) 
\int \d^2 z \, 
{\delta\rho}{\delta \lambda}
-\frac{d\Lambda^2}{2\blambda} \!\!\int \d^2 z \, \brho 
\Big(\frac{\delta \lambda}{\blambda}\Big)^2
+\frac{(26-d)}{96\pi } \int \d^2  z \, \Big(\frac{\partial_a \delta \rho}{\brho}\Big)^2
 \non &&-\frac{d}{24\pi} \int \d^2 z \, \Big(\frac{\partial_a \delta \rho}{\brho}\Big) 
\Big(\frac{\partial_a \delta \lambda}{\blambda}\Big)
+\frac{ d }{32 \pi } 
\int \frac{\d^2 p }{(2\pi)^2}\,\Big(\frac{\delta \lambda(p)}{\blambda}\Big)
\Big(\frac{\delta\lambda(-p)}{\blambda}\Big)\;
p^2\log\Big(\frac{1}{ \epsilon p^2}\Big)
\label{S2}
\eea
for a certain $\epsilon\sim 1/\Lambda^2$.
Notice the last term on the right-hand side is normal  (and therefore regularization dependent) rather than anomalous as the third and fourth terms.

{}From \eq{S2} for the effective action 
to the second order  in fluctuations we find the  quadratic form
\be
\delta S_2=\int \d^2z \Big[\delta \left(\log\rho\right)
A_{\rho\rho}\,\delta \left(\log\rho\right) \non +
2\,\delta \left(\log\rho \right)A_{\rho\lambda} \,
\delta \left(\log\lambda\right) 
+\delta \left(\log \lambda\right)A_{\lambda\lambda} \,\delta \left(\log\lambda \right)
 \Big]
\label{191}
\ee
with
\be
A_{ij}(p) = \left[ \begin{array}{cc} 
\frac {(26-d) p^2}{96\pi}& 
 -\frac12\left(K_0-\frac{d\Lambda^2 }{2\blambda^2}\right)\brho(p) \blambda
-\frac {d p^2}{48\pi}\\ 
-\frac12\left(K_0-\frac{d\Lambda^2 }{2\blambda^2}\right)\brho(p) \blambda
-\frac {d p^2}{48\pi}
&  -A(p)
\end{array}
\right],
\label{MMMM}
\ee
\end{widetext}
where
\be
A(p)=  \frac{d \Lambda^2\bar \rho(p)} {2\blambda} +\frac {d p^2}{32\pi}\log (\epsilon p^2) 
\label{AAAA}
\ee
is always positive.

Since $\delta \lambda(z)$ is
pure imaginary, i.e.\ $\delta \lambda (- p) = - \delta\lambda^*(p)$, we find for the determinant associated with the matrix 
 in \eq{MMMM} 
\be
D= \left[\frac12\left(K_0-\frac{d\Lambda^2}{2\blambda^2}\right) \brho \blambda+
\frac {d p^2}{48\pi}\right]^2+ \frac {(26-d) p^2}{96\pi} A
\label{D}
\ee
and the propagators corresponding to the action \rf{191} are given by 
\beq\label{xj3}
\LA \phi^*_i(p) \phi_j(p) \RA = \frac{A_{ij}}{D},
\quad \phi_i =\left[\delta \left(\log \rho\right)
, \; 
\delta \left(\log\lambda\right)\right]. 
\eeq

The situation with stability of fluctuations is just the same as described in \cite{AM17}:
they are unconditionally stable for $2<d<26$. For $d>26$ they are stable in the regularized case, 
where $\Lambda $ is large but finite, because the first term
on the right-hand side of \eq{D} then dominates.
In the scaling regime \rf{sca}, 
where $K_R$ is finite as $\Lambda\to\infty$,
we have the situation where the first term on the right-hand side of \eq{D} is
finite after the renormalization of $\brho$, so the second term dominates.
The action \rf{191} thus becomes unstable for $d>26$ in the Lilliputian scaling limit.

\section{Conclusion}

We have computed scattering amplitudes of the regularized bosonic Nambu-Goto
string in the mean-field approximation, disregarding fluctuations
of the Lagrange multiplier and an independent metric about their 
mean values. We have
recovered  the Regge behavior of the amplitudes in the Regge regime 
with the usual linear Regge trajectory in 
space-time dimensions $d>2$.
We have considered the Lilliputian scaling limit,  previously introduced \cite{AM15b}
to recover the results of canonical quantization for the spectrum the Dirichlet string,
and showed that it is the one where the slop of the Regge trajectory scales to a finite
value.
We have also demonstrated that the effective action indeed has a stable minimum given 
by the mean field and fluctuations about the mean
values of $\lambda^{ab}$ and $\rho$ increase the effective action for $2<d<26$.

The fact that the mean-field approximation reproduces canonical quantization in $d=26$ is
not surprising and is well understood. However, canonical quantization is not consistent in $2<d<26$,
where the effective action depends  nonlinearly on  the world sheet metric $\rho$. 
An  open interesting question is as to whether or not corrections to the mean field
will change the above formulas in $2<d<26$.

\subsection*{Acknowledgments}

The authors acknowledge  support by  the ERC Advanced
grant 291092, ``Exploring the Quantum Universe'' (EQU).
Y.~M.\ thanks the Theoretical Particle Physics and Cosmology group 
at the Niels Bohr Institute for the hospitality. 


\end{document}